\documentclass{PoS}
\pdfoutput=1 
\title{Interactive Visualization Package for 4D Lattice Field Theories}

\ShortTitle{Interactive Visualization Package for 4D LFTs}

\author{Ivan Hip
        \thanks{Work supported by the Croatian Ministry of Science, Education and Sports (project No. 0160013).}\\
       University of Zagreb, Faculty of Geotechnical Engineering\\
       Hallerova aleja 7, 42000 Vara\v{z}din, Croatia\\
       E-mail: \email{ivan.hip@gmail.com}}

\abstract{Recent interest in exploring local vacuum structure of QCD through the properties of the eigenmodes of the lattice Dirac operators rises again the challenge to visualize four-dimensional objects and structures which appear in lattice field theories. In spite of complex and powerful commercial visualization software packages on the market, there are reasons to develop Interactive Visualization Package (IVP). We believe that an apprehension of the complex structures is possible only through the interactive approach, with the user being able to  manipulate data representations and slices through the lattice in real-time. Further insight should also be gained by an interactive parallel examination of different physical quantities, e.g. eigenmode density with topological charge or action densities. Finally, thanks to constantly falling hardware prices, IVP makes it possible to use almost any Linux PC as a visualization tool for research in lattice field theory.}

\FullConference{The XXV International Symposium on Lattice Field Theory\\
                July 30 - August 4 2007\\
                Regensburg, Germany}

\begin{document}

\begin{figure}
\includegraphics[width=\textwidth]{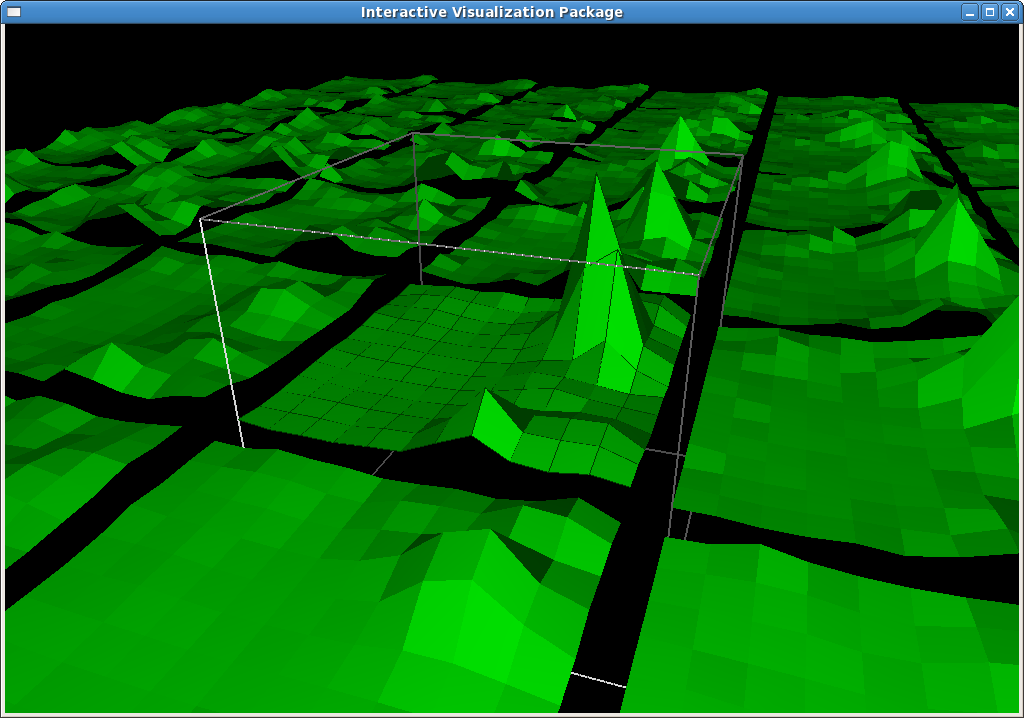}
\caption{Plot of the eigenmode density:
An array of surfaces, each representing data values on 
a two-dimensional slice through the lattice when two of the coordinates (e.g. $z$ and $t$)
are kept fixed, is shown.
The actual  $(z, t)$ position in the lattice is indicated by the white wireframe
cube. The position can be interactively changed by pressing the arrow keys.}
\label{peak}
\end{figure}

\begin{figure}
\includegraphics[width=\textwidth]{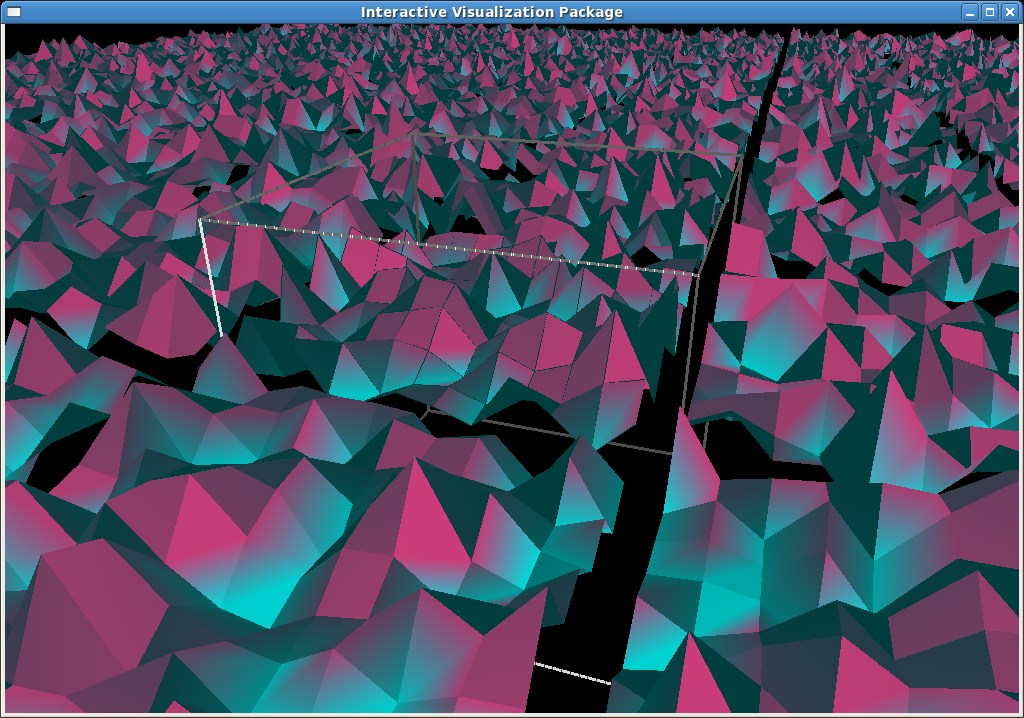}
\caption{Gauge action density - a sea of seemingly random fluctuations. Color is function of density.}
\label{rawgauge}
\end{figure}

\begin{figure}
\includegraphics[width=\textwidth]{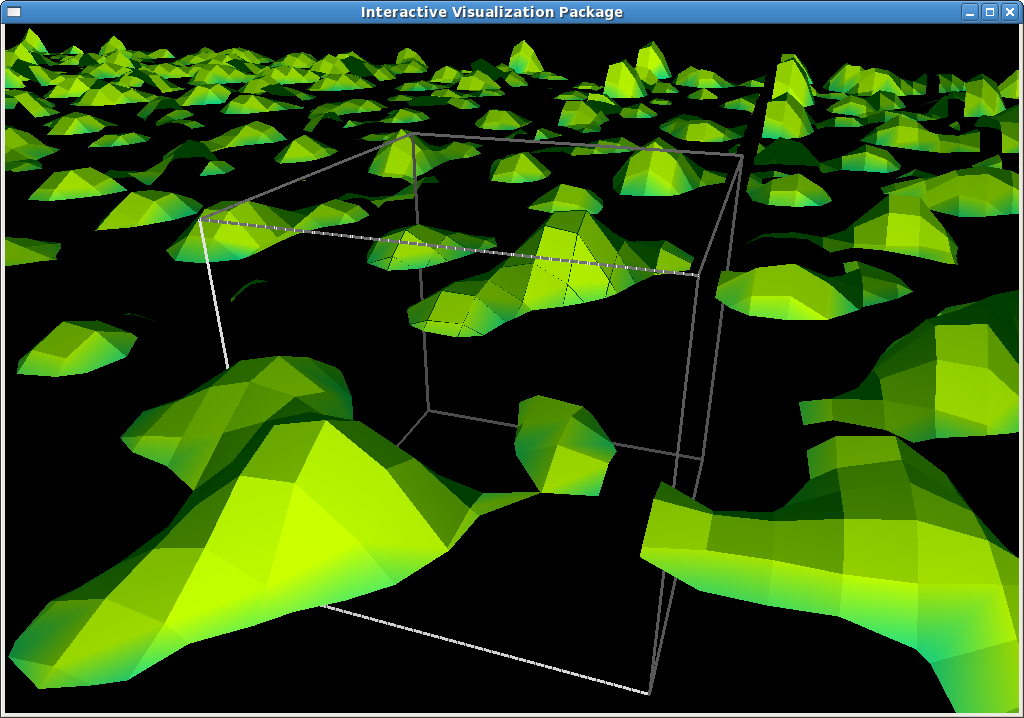}
\caption{After clipping by a horizontal plane only the peaks remain visible. The position of the clipping plane
can be adjusted interactively.}
\label{clipping}
\end{figure}

\begin{figure}
\includegraphics[width=\textwidth]{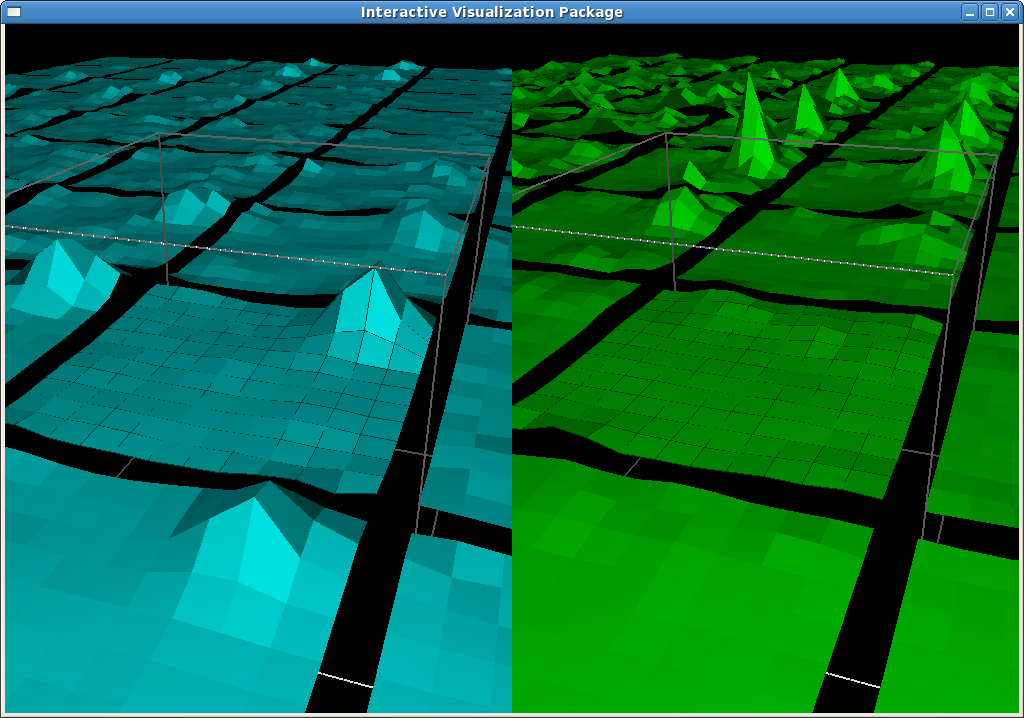}
\caption{Comparison of peak positions for two eigenmodes.}
\label{vsplit}
\end{figure}

\begin{figure}
\includegraphics[width=\textwidth]{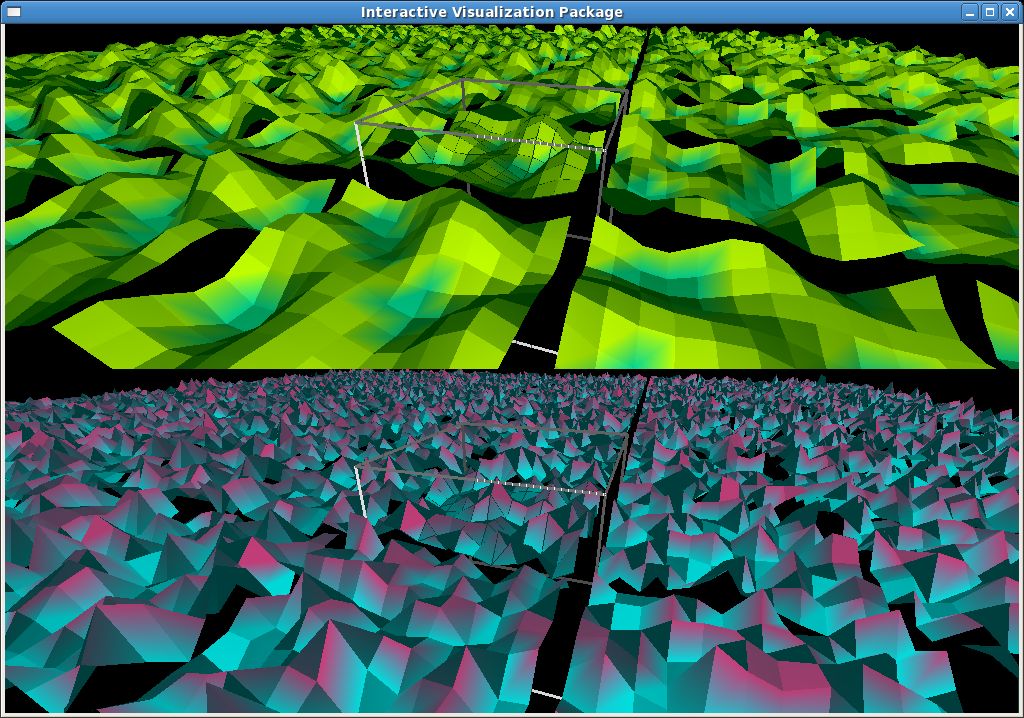}
\caption{Direct comparison of the filtered gauge action density (top) with the raw gauge action density (bottom).}
\label{fig5}
\end{figure}

\begin{figure}[t]
\includegraphics[width=\textwidth]{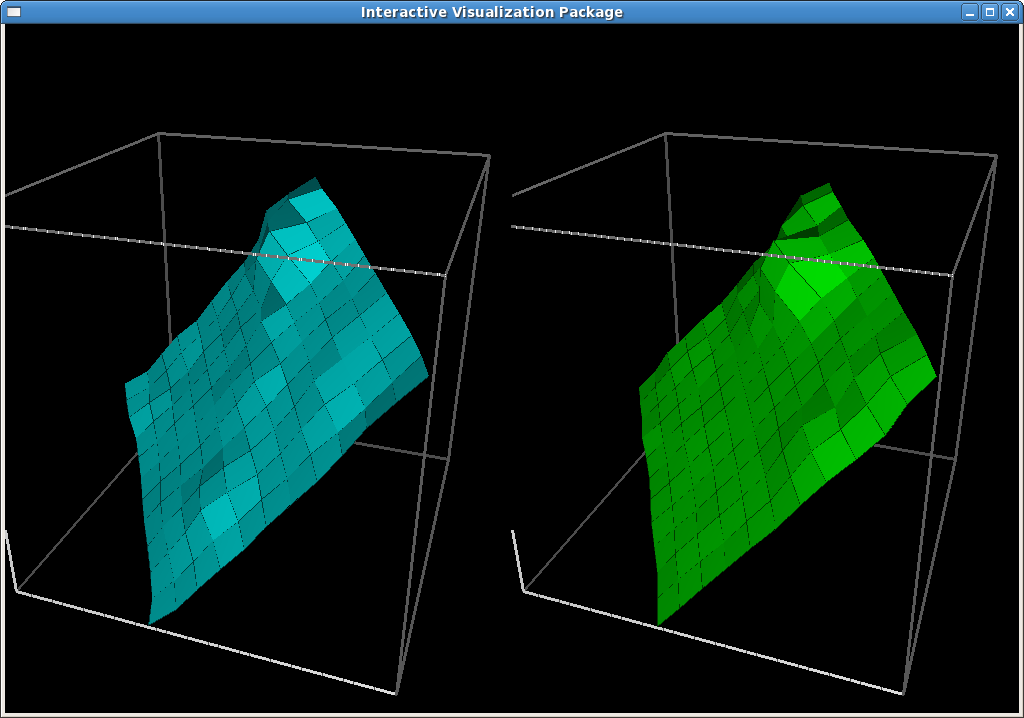}
\caption{Selected slices can be rotated side by side in real-time.}
\label{rot3}
\end{figure}

\section{Introduction}

The momentum given to the exploration of local vacuum structure of QCD through
the properties of the eigenmodes of the lattice Dirac operators by Horvath et al.
\cite{Horvath:2001ir} does not decrease. 
In recent work Gattringer, Ilgenfritz and their collaborators
 \cite{Gattringer:2006wq, Bruckmann:2006wf, Ilgenfritz:2007xu} use eigenmodes to
compare filtering methods and properties of topological excitations.  
Anna Hasenfratz et al. \cite{Hasenfratz:2007iv} draw the attention to the role played by the
localized eigenmodes.
All these current investigations rose again the challenge to visualize and compare four-dimensional
objects and structures which appear in lattice field theories.

There was a lot of effort to visualize topological objects on the lattice with the aim to study
their interplay and to gain better insight into the cooling process.
In the second half of the 1990s  the Viennese group \cite{Feurstein:1996gq}
paid special attention to the visualization of these phenomena.
Visually very effective and well-known are the visualizations by 
Leinweber \cite{Leinweber:1999cw} and some of the most sophisticated
visualization schemes were proposed by Gutbrod \cite{Gutbrod:2000sp}.
Although all of these visualization schemes use animation, none is really
interactive in the immersive virtual reality sense which is implemented by
Interactive Visualisation Package (IVP) as it is presented in this paper.
IVP allows the user to move interactively through the data, to compare the data
by showing data sets side by side or to identify peaks and possible other
artifacts by using clipping planes and coloring effects.

\section{Concept}

It is a lasting challenge how to present a four-dimensional data set
to be easily percepted by the human brain. Of course, one of the four 
dimensions can be treated as time and we are left with a problem 
to express the changes of a three-dimensional function by animation.
But still, there is again no simple way to present values of the function defined
in three-dimensional space on a two-dimensional computer screen because
the interior points are inevitably screened. The usual way out is to render only
some subsets of the points, e.g. isosurfaces, where the function has identical
values.

A clean way to present every function value in every point is 
to render just two-dimensional slices with function
values shown on the third axis. In the case of four-dimensional lattice
that means to fix two of the lattice coordinates (e.g. $z$ and $t$) and 
to handle the values of the lattice function for every $(x, y)$ coordinate pair as
height. Four such values (vertices) define quadrilaterals
(four sided polygons) which are easy to plot and they define
two-dimensional surface which presents the function behavior on the
lattice slice under consideration.
However, we are left with $L^2$ (for simplicity we assume a cubic
lattice with size $L$) such surfaces and they often look rather similar
to each other, which makes it tedious to examine them one by one.
An obvious way to get some insight into the whole lattice
volume is to plot $L^2$ such independent surfaces in a rectangular array.
But, individual surfaces become too small and there is no real insight until
separate plots are joined together to form a common virtual plane.
Perspective projection of this plane enhanced by color and lighting effects
can be rendered on a computer screen and by enabling interactive walking
through this scenery we get a new essential quality:
instead of a hardly imaginable four-dimensional world, we immerse in
a landscape with mountains and valleys familiar to
our eyes and everyday experience. In such a situation the superb abilities
of human brain for orientation in space and pattern recognition
can come to their full advantage. Now it is much easier to identify some 
characteristic shapes and to remember the position in the landscape
(i.e. lattice) where such artifacts have been observed.

Fig. 1 illustrates the concept: just by pressing the arrow keys one can
fly above an array of two-dimensional slices. Moving up or down means
to change the $z$-coordinate, and moving left or right $t$-coordinate
of the slices in the four-dimensional lattice.

The color can be coded dependent on the data values to emphasize peaks
or lakes (Fig. 2). The position of the color threshold can be changed interactively.
Another, but similar, visual aid is implemented by a clipping plane. By rising
a clipping plane, the surfaces bellow the plane become invisible, and only
the peaks above some threshold are left in the scenery (Fig. 3). Again,
the position of the clipping plane can be adjusted interactively.

Further insight is provided by an interactive simultaneous examination
of two different configurations, eigenmodes or even two different physical quantities
(e.g. eigenmode density vs. topological charge or action densities) at the same
position on the lattice.
IVP reads two four-dimensional data sets and allows to switch between them
by pressing '1' or '2', or even to split the screen vertically (Fig. 4) or
horizontally (Fig. 5) --- just by pressing 'v' or 'h' on the keyboard. 

Finally, if one wishes to compare two slices in more detail, one can apply
simultaneous parallel rotation of the slices illustrated in Fig. 6.

\section{Coding}

IVP relies on technologies used for implementation of modern computer games
with complex real-time three-dimensional graphics rendering.
Thanks to the power of the OpenGL graphics library enhanced with OpenGL Utility Libraries
(GLUT) only a few hundred lines of standard C code were enough to provide 
optimal functionality and amazing visual effects.
OpenGL is one of the two (another is DirectX) computer graphics industry
standards which is widely supported by graphics cards manufacturers.
Many OpenGL routines are performed directly by graphics hardware and
that, combined with the coding in C programming language, makes IVP fast
and efficient on almost any PC or even laptop.
For a simple introduction to OpenGL and GLUT libraries see \cite{openGL}.

IVP was successfully tested on Fedora and Debian Linux and should be easily ported to any
platform with OpenGL and GLUT libraries available.
If you are interested to test or apply IVP, please write to \texttt{ivan.hip@gmail.com}.

\acknowledgments
I would like to thank Christof Gattringer and Christian B. Lang for providing me with gauge
configurations and eigenmodes, to Kre\v{s}imir Kumeri\v{c}ki who tested the IVP on Debian
Linux and to Mario Zidar who helped with the design of the poster.

\end{document}